\def\year{2018}\relax

\documentclass[letterpaper]{article} 
\usepackage{aaai18}  
\usepackage{times}  
\usepackage{helvet}  
\usepackage{courier}  
\usepackage{url}  
\usepackage{graphicx}  
\usepackage{amsmath}
\usepackage{booktabs}
\usepackage{paralist}

\usepackage{pgfplotstable}
\pgfplotsset{compat=1.13}

\newcommand{\citet}[1]{\citeauthor{#1} (\citeyear{#1})}

\begin{document} 

\title{The Effect of Extremist Violence on Hateful Speech Online}
\author{ 
Alexandra Olteanu \\
       IBM Research \\
       alexandra.olteanu@ibm.com \\
\And 
Carlos Castillo \\
       UPF, Barcelona \\
       chato@acm.org  \\
\And
Jeremy Boy \\
	   UN Global Pulse\\
       jeremy@unglobalpulse.org  \\
\And
Kush R. Varshney \\
	   IBM Research \\
       krvarshn@us.ibm.com  \\
}
\maketitle

\begin{abstract}
User-generated content online is shaped by many factors, including endogenous elements such as platform affordances and norms, as well as exogenous elements, in particular significant events. 
These impact what users say, how they say it, and when they say it. 
In this paper, we focus on quantifying the impact of violent events on various types of hate speech, from offensive and derogatory to intimidation and explicit calls for violence.
We anchor this study in a series of attacks involving Arabs and Muslims as perpetrators or victims, occurring in Western countries, 
that have been covered extensively by news media.
These attacks have fueled intense policy debates around immigration in various fora, including online media, which have been marred by racist prejudice and hateful speech.
The focus of our research is to model the effect of the attacks on the volume and type of hateful speech on two social media platforms, Twitter and Reddit. 
Among other findings, we observe that extremist violence tends to lead to an increase in online hate speech, particularly on messages directly advocating violence.
Our research has implications for the way in which hate speech online is monitored and suggests ways in which it could be fought.
\end{abstract}

\section{Introduction}

%
Hate speech is pervasive and can have serious consequences. 
According to a Special Rapporteur to the UN Humans Rights Council, failure to monitor and react to hate speech in a timely manner can reinforce the subordination of targeted minorities, making them ``vulnerable to attacks, but also influencing majority populations and potentially making them more indifferent to the various manifestations of such hatred''~\cite{izsak2015hate}.   
At the individual level, people targeted by hate speech describe ``living in fear'' of the possibility that online threats may materialize in the ``real world''~\cite{awan2015we}. 
At the level of society, hate speech in social media has contributed to fuel tensions among communities, in some cases leading to violent clashes~\cite{izsak2015hate}.

%
Following one of the most severe humanitarian crises in recent history, Europe has seen a high immigration influx, including Syrian, Afghan, and Iraqi refugees.\footnote{World Economic Forum (Dec. 2016) ``Europe's refugee and migrant crisis'' \url{https://www.weforum.org/agenda/2016/12/europes-refugee-and-migrant-crisis-in-2016-in-numbers}} 
In the same period, several deadly terror attacks have occured in Western nations~\cite{reuters-attacks-Europe,global-terrorism-database},
leading to an increasingly alarming anti-Muslim rhetoric~\cite{tellmama} by right-wing populist movements~\cite{Greven2016} and right-leaning media outlets~\cite{the-independant-Sun}, often conflating refugees and Muslims with Islamic fanatics~\cite{diene2006muslims}.  
This rhetoric has also gained adoption online~\cite{UNHCR-UNGP}, prompting governmental agencies\footnote{BBC News (Sep. 2017) ``Social media warned to crack down on hate speech'' \url{http://bbc.com/news/technology-41442958}} and NGOs to call on social media platforms to step up their efforts to address the problem of hate speech~\cite{the-independant-hate-crimes,tellmama}.   
The concern is that the increase in hateful narratives online led to an upsurge in hate crimes targeting Muslim communities~\cite{the-independant-hate-crimes}.  
Insights into how online expressions of hate thrive and spread can help stakeholders' efforts to de-escalate existing tensions~\cite{burnap2014hate}. 

\begin{figure*}
  \centering
    \includegraphics[width=0.82\textwidth]{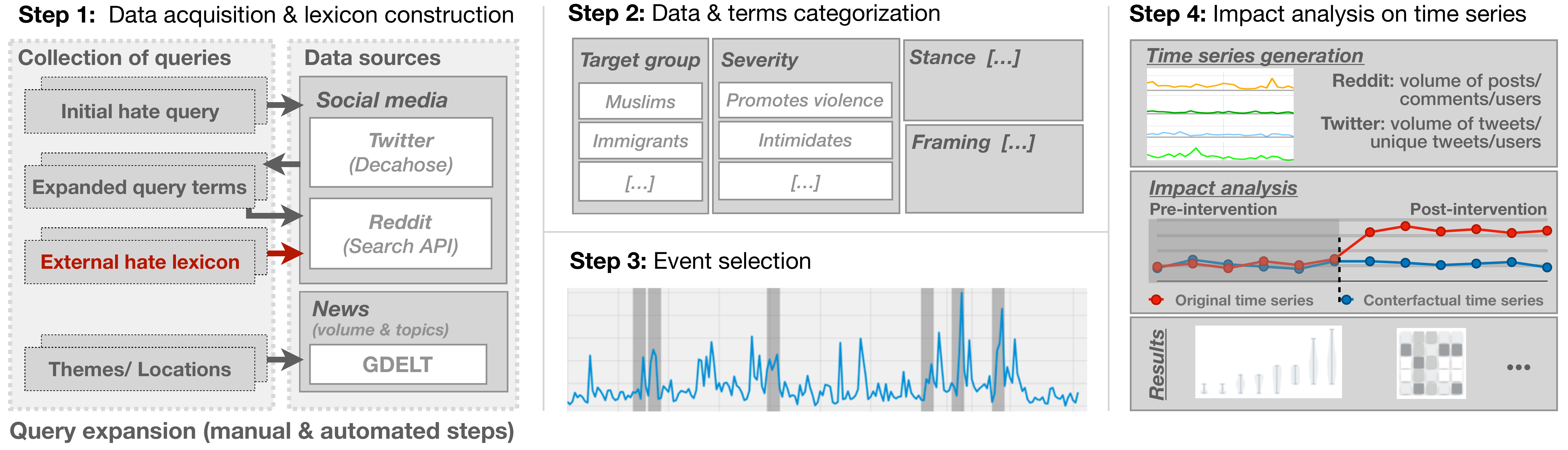}
     \caption{Steps in our analysis framework: 
     \inparaenum[(1)]{
     \item data acquisition and lexicon creation, \S\ref{sec:data}; 
     \item data categorization, \S\ref{sec:characterizing};
     \item events selection, \S\ref{subsec:events};
     \item impact analysis on time series of hate related terms, \S\ref{sec:method} and results, \S\ref{subsec:empirical-results}.}
     }
      \label{analysis-framework}
\end{figure*}

In this paper, we explore {\em how hate speech targeting specific groups on social media is affected by external events}. 
Anchoring our analysis in a series of Islamophobic and Islamist terrorism attacks in Western countries, we study their impact on the prevalence and type of hate and counter-hate speech targeting Muslims and Islam on two different social media platforms: Twitter and Reddit. 

\noindent {\bf Our contribution.  }
We conduct a quantitative exploration of the causal impact of specific types of external, non-platform specific events on social media phenomena. 
For this, we create a lexicon of hate speech terms, as well as a collection of 150M+ hate speech messages, propose a multidimensional taxonomy of online hate speech, and show that a causal inference approach contributes to understanding how online hate speech fluctuates.
Among our findings, we observe that extremist violence attacks tend to lead to more messages directly advocating violence, demonstrating that concerns about a positive feedback loop between violence ``offline'' and hate speech online are, unfortunately, well-founded.

\subsubsection*{Paper Outline and Methodology Overview. } 
Outlined in Figure~\ref{analysis-framework}, our approach consists of several steps:

\noindent \textbf{Step 1: }
We create a longitudinal collection of hate speech messages in social media from Twitter and Reddit that covers a period of 19 months. 
This collection is based on a series of keywords that are obtained through an iterative expansion of known hate speech terms (\S\ref{sec:data}).

\noindent \textbf{Step 2: }
We categorize the data along four dimensions:  
\begin{inparaenum}[1)]
\item the group each message refers to, 
\item the attitude of speakers,
\item the severity of hateful expressions, particularly whether they advocate violence, and
\item the framing of content 
\end{inparaenum}
(\S\ref{sec:characterizing}).

\noindent \textbf{Step 3: }
We select 13 extremist attacks involving Arabs and Muslims as perpetrators or victims, like the Berlin Christmas market attack on Dec. 2016, perpetrated by a follower of jihadist group ISIL, or the Quebec City mosque shooting on Jan. 2017, 
by a far-right white nationalist (Table~\ref{tab:external_events}).

\noindent \textbf{Step 4: }
As evaluating the effect of such attacks on various slices of social media is a causal question, we frame it as measuring the impact of an intervention (event) on a time series (temporal evolution of speech).  
Following techniques for causal inference on time series~\cite{brodersen2015inferring},  we estimate an event's impact on various types of hate and counter-hate speech by comparing the behavior of corresponding time series after an event, with counterfactual predictions of this behavior had no event taken place (\S\ref{sec:method}).  

The last sections present (\S\ref{sec:analysis}) and discuss (\S\ref{sec:conclusion}) our results.

\section{Background \& Prior Work}
\label{sec:background}

We are interested in the relation between online hate speech and events.
To ground our study, we first review work defining hate and counter-hate speech. 
Given our focus on anti-Muslim rhetoric in the context of extremist violence, we outline previous works on hate speech after terror attacks, and studies of hateful narratives targeting Muslims. 
We also cover observational studies on social media, particularly those focusing on harmful speech online. 

\subsection{Hate Speech Online and Offline}
\label{hate-context}

\noindent {\bf Defining hate speech. }
While hate speech is codified by law in many countries, the definitions vary across jurisdictions~\cite{sellars2016defining}. 
As we study hate speech directed at followers of a religion (Islam) that is often conflated with a particular ethnicity and culture, we use the definition by the Council of Europe covering ``all forms of expression which spread, incite, promote or justify racial hatred, xenophobia, anti-Semitism or other forms of hatred based on intolerance, including: intolerance expressed by aggressive nationalism and ethnocentrism, discrimination and hostility against minorities, migrants and people of immigrant origin.''~\cite{ce1997hate}. 
We extend this definition to further consider as hate speech ``animosity or disparagement of an individual or a group on account of a group characteristic such as race, color, national origin, sex, disability, religion, or sexual orientation''~\cite{nockleby2000hate}, allowing us to juxtapose hateful speech directed at Muslims with other groups (e.g., LGBTQ, immigrants).

\noindent {\bf Counter-hate speech. }
Censoring hate speech may clash with legal protections on free speech rights. 
Partially due to this tension, the position of international agencies like UNESCO is that ``the free flow of information should always be the norm. Counter-speech is generally preferable to suppression of speech"~\cite{gagliardone2015countering}. 
Thus, it is important not only to study hate speech, but also to contrast it with counter-speech efforts---a rare juxtaposition in social media research~\cite{benesch2016counterspeech,magdy2016isisisnotislam}. 
\citet{magdy2016isisisnotislam} estimate that a majority of Islam and Muslim related tweets posted in reaction to the 2015 terrorist attacks in Paris stood in their defense; an observation also made by \citet{UNHCR-UNGP} following the 2016 terrorist attack in Berlin, and supported by our own results (\S\ref{subsec:empirical-results}).

\noindent {\bf Hate speech and violent events. } 
The prevalence and severity of hate hate speech and crimes tends to increase after ``trigger'' events, which can be local, national, or international, often galvanizing ``tensions and sentiments against the suspected perpetrators and groups associated with them"~\cite{awan2015we}.    
For instance, \citet{benesch2016counterspeech} found extensive hate and counter-hate speech after events that triggered widespread emotional response like the Baltimore protests, the U.S. Supreme Court decision on same-sex marriage, and the Paris attacks during 2015;  
while \citet{faris2016understanding} found spikes in online harmful speech to be linked to political events. 
While these studies are related to ours, they focus on content posted during specific events, or on correlating changes in patterns (e.g., spikes) with events' occurrence. 
We focus on broader patterns, aiming to quantify changes across types of events and types of content by applying causal inference techniques.

\noindent {\bf Islamophobia. }  
The conflation of Muslims and Islam with terrorism---particularly developed after September 11, 2001---is a key factor behind the increase in Islamophobic attitudes~\cite{diene2006muslims}. 
A significant increase in anti-Muslim hate crimes was observed after terrorist attacks by individuals that identify as ``Muslim or  acting in the name of Islam," with those having  a ``visible Muslim identity'' being the most vulnerable to hostility, including online and offline intimidation, abuse and  threats of violence~\cite{awan2015we}.

\subsection{Observational Studies Using Social Media}
\label{hate-online}

\noindent {\bf Hate speech on online social platforms. }  
While social media platforms provide tools to meet new people, maintain relationships, promote ideas, 
and promote oneself; 
they have also opened up new avenues for harassment based on physical appearance, race, ethnicity, and gender~\cite{duggan2017harassment}.
This has led to efforts to detect, understand, and quantify such harmful speech online, with goals such as modeling socially deviant behavior~\cite{cheng2017anyone}, building better content filtering and moderation tools~\cite{matias2015reporting}, and informing policy makers~\cite{faris2016understanding}. 

The main categorization criteria for online hate speech has been based on 
the group being targeted (e.g., ``black people,'' ``fat people''), 
the basis for hate (e.g., race, religion)~\cite{silva2016analyzing,saleem2016web},
and the speech severity~\cite{davidson2017automated}.   
For instance,~\citet{silva2016analyzing} found ``soft" targets like ``fat people" to be among the top target groups;  
yet, these groups are often not included in the documentation of offline hate crimes.
\citet{davidson2017automated} further discuss challenges in distinguishing between hate speech and other types of offensive speech.

\noindent {\bf Observational methods applied to social data. }   
Recent studies show that quasi-causal methods can be applied to social media data to e.g., distill the outcomes of a given situation \cite{olteanu2017distilling}, measure the impact of an intervention~\cite{chandrasekharan2018you}, or estimate the effect of online social support~\cite{cunha2017warm}. 
The application of these methods to social data, including propensity score matching~\cite{de2016discovering}, difference-in-differences~\cite{chandrasekharan2018you}, and instrumental variables~\cite{zhang2016understanding}, was found to reduce confounding biases.

\citet{chandrasekharan2018you}'s work is closest to ours, as it employs techniques from the causal inference literature to quantify the impact of an intervention on hateful behavior on Reddit.  
Yet, the intervention they study is platform-specific--a ban on an existing community on Reddit--whereas we look at the impact of external (non-platform specific) events on both Reddit and Twitter.  
Our focus is on the overall prevalence of hate speech, rather than on the behavior of given groups of users, and we measure the effect of given interventions (events) on various types of hate speech (\S\ref{sec:characterizing}).

\noindent {\bf Operationalization of hate speech on social media. } \label{subsec:hate-operationalization} 
Due to lack of consensus on what constitutes hate speech and the challenges in operationalizing existing definitions at the scale of current online platforms, prior work has used a mix of manual and automated term selection strategies to identify terms that are likely to occur in hateful texts~\cite{chandrasekharan2018you,davidson2017automated}.
While focusing on speech targeting Muslims and Islam, we similarly combine existing lexicons with terms obtained through a combination of manual and automated steps (\S\ref{subsec:query-construction}).

\section{Data Collection}
\label{sec:data}

Our goal is to characterize and measure online hate speech targeting Muslims and Islam in reaction to major Islamist terror attacks and Islamophobic attacks perpetrated in Western countries. 
Here, we describe our data collection process, which attempts to be inclusive (high-recall) and hence uses a broad definition of hate and counter-hate speech.
We iteratively expand an initial query of keywords related to relevant items by identifying new keywords in the retrieved messages. 
Our base datasets contain messages from Twitter and Reddit, and a collection of news articles; these are not associated to any particular event, but cover messages potentially related to hate and counter-hate speech over a period of 19 months: from January 1, 2016 to August 1, 2017.

\subsection{Data Sources}\label{subsec:data-sources}

{\bf Twitter} (\url{https://twitter.com/})
is one of the largest microblogging platforms used by hundreds of millions every month.
To collect Twitter messages (``tweets'') we use an archive representing 10\% of the entire public stream, known as the ``Decahose.'' 

\noindent {\bf Reddit} (\url{https://reddit.com/})
is a large social news aggregation platform used by millions every month.
Users submit ``posts'' and ``comments'' that gain or lose visibility according to up- and down-votes. 
We collect posts through Reddit's Search API\footnote{Using the PRAW library: \url{https://praw.readthedocs.io/}.} (comments are not searchable via~this API), retaining all comments to posts matching our queries.

\noindent {\bf News.}
Finally, we collect news articles from GDELT (Global Data on Events, Location, and Tone, \url{http://gdeltproject.org/}), the largest online catalog of global news events.
We use these data as exogenous variables when modeling social media time series before and after a given event.

\subsection{Query Construction}\label{subsec:query-construction}

We collected data using keyword queries, a sampling method applicable to both Twitter's and Reddit's APIs.
As our goal was to create a high-recall collection, our sampling procedure consists in formulating an initial query (bootstrapping), followed by an expansion of that query. This method is known to improve the coverage of social media data~\cite{olteanu2014crisislex,davidson2017automated}. 

\noindent {\bf Query Bootstrapping.}\label{concept:query-bootstrapping}
We bootstrapped our query selection with an initial list of terms (keywords and hashtags) used in social media campaigns related to anti-Muslim hate and counter-hate speech.  
This list was assembled retrospectively (as was the rest of our data) using
\begin{inparaenum}[(i)]
\item news articles and blog posts discussing social media usage during hate and counter-hate campaigns,\footnote{E.g.,~\url{http://huffingtonpost.com/entry/anti-muslim-facebook-twitter-posts-myths-wajahat-ali_us_57f55bb1e4b002a7312084eb}~or~\url{http://muslimgirl.com/29612/10-hashtags-literally-changed-way-muslims-clap-back/}}
\item resources from NGOs or governmental agencies tracking or analyzing hate speech on social media \cite{awan2015we,UNHCR-UNGP}, and 
\item research articles~\cite{magdy2016isisisnotislam}. 
\end{inparaenum}
Selected terms were individually validated by manually searching for them on both Twitter and Reddit. 
Additional co-occurring terms found in this process were added to the list. 
This step resulted in a list of $91$ terms, including ```\#f***quran,'' ``\#nosharia,'' ``ban islam,''  and ``kill all muslims.''

\noindent {\bf Query Expansion. }  
We then employed a query expansion heuristic to identify further terms that may appear in messages expressing hate or counter-hate towards different groups, including, but not limited to, Arabs and Muslims.
The heuristic considers terms frequently appearing in social media messages matched by the terms in our initial list. 
To obtain a high-recall collection, we considered any new term that may constitute hate or counter-hate speech, using an inclusive, broad definition inspired by~\citet{silva2016analyzing} and \citet{chatzakou2017mean}, and expanded to also cover commentary and counter-hate speech elements. 
We recorded all terms related to \emph{speech that could be perceived as offensive, derogatory, or in any way harmful, and that is motivated, in whole or in a part, by someone's bias against an aspect of a group of people, or related to commentary about such speech by others, or related to speech that aims to counter any type of speech that this definition covers.}

This expansion was independently done in two iterations for both Twitter and Reddit.
First, one of the authors did an annotation pass to identify new query terms.  
Second, as we favored recall, at least one other author did an additional annotation pass over the terms rejected by the first annotator.

\noindent {\bf External lexicon. }
To further expand our list of query terms, we added terms from a lexicon built using HateBase,\footnote{HateBase: \url{https://www.hatebase.org/}} a website that compiles phrases submitted and tagged by internet users as constituting hate speech.
Given that only an estimated 5\% of messages containing HateBase terms were actually identified as hateful;  
instead of directly using these terms, we used 163 unique terms extracted from Twitter messages containing HateBase terms and manually annotated as hateful or offensive by~\citet{davidson2017automated}.\footnote{Given that our queries represent a conjunction of the words in each term and that for Reddit articles are ignored, we pre-process most terms to remove the article {\em a} and remove duplicates.}

\subsection{Data Acquisition}
\label{data-acquisition}

Table~\ref{tbl:datasets} presents a summary of the data we acquired.
 
\begin{table}[tb]
\scriptsize
\centering
\begin{tabular}{lrrrr} \toprule
{\bf Source  }
 & \multicolumn{1}{c}{\bf Terms}
 & \multicolumn{1}{c}{\bf Messages}
 & \multicolumn{1}{c}{\bf Users}
 & \multicolumn{1}{c}{\bf URLs} \\ \midrule
{\it Twitter}	& 825    
& 107M & 26M  & -- \\ 
{\it Reddit}	& 1,257   
& 45M    & 3.3M  & -- \\ 
{\it News}	&  --      
&  --            & --   & 3.3 M \\
\bottomrule
\end{tabular}
\caption{Summary of the final data collection.} 
\label{tbl:datasets}
\end{table}

\noindent {\bf Acquiring Twitter data.}
We first queried the bootstrap terms, and retrieved 958K messages posted by 413K users.
We then expanded the query by manually annotating 2088 terms that appeared more frequently than an arbitrary threshold (ranging from 75 for tri-grams to 300 for uni-grams, which are typically less precise than tri-grams and noisier at lower frequencies), after removing stopwords using the Python NLTK package.
We found an extra 612 terms. 
We queried these terms, growing our collection by 55M tweets posted by 12.5M users.  
The resulting dataset contains on average 4.5M tweets per month. 
Since we used the Twitter Decahose (a 10\% sample of all Twitter content), we estimate this collection is in fact representative of a larger set of roughly 45M tweets per month. 
Finally, we retrieved tweets matching the 163 external hate terms (based on HateBase), resulting in an additional 51.6M tweets by 13.7M users. 
Altogether, we collected over 1TB of raw Twitter data.

\noindent {\bf Acquiring Reddit data. }
We again began by querying the bootstrap terms, and retrieved 3K posts with 140K comments written by 49K users. 
We then expanded the query by selecting high-frequency terms (thresholds ranging from 50 to 300 as these data were sparser than Twitter) across all posts and comments, and manually annotating them. 
Given that the Reddit Search API normalizes terms before running a query, we did not keep different inflections of the same terms.
We annotated 4272 terms, and found 1002 related to hate and counter-hate speech. 
We queried these terms, and retrieved an extra 300K posts with 41M comments written by 3.1M users.
Finally, we queried the external hate terms. 
Altogether, we collected 337K posts with 45M comments written by roughly 3.3M users.

\noindent {\bf Acquiring news data.} 
We used GDELT's Global Knowledge Graph (GKG), as it provides the list of news articles covering each event in their database.  
This allowed us to compute the overall volume of news per day, amounting to over 130M URLs over our 19 months period of interest.

\begin{table*}[ht]
\scriptsize
\begin{tabular}{p{0.6in}p{2.6in}|p{0.6in}p{2.6in}@{}}\toprule
Theme & Paraphrased message/comment(s) &  
Theme & Paraphrased message/comment(s) \\\midrule
Blames \mbox{Muslims} 
&
\begin{minipage}[t]{2.6in}
``{\em Ban Muslims, and you won't have Islamic terrorism}" (T)\\
``{\em Islam is the problem and everyone knows this}" (R) 
\end{minipage}
&
Defends \mbox{Muslims}
&
\begin{minipage}[t]{2.6in}
``{\em killing innocent people is not Islam, there were Muslims at that concert as well}" (T) ``{\em \#IllRideWithYou indicates one should not be scared to be a Muslim. One should be scared to be a racist}'' (T)
\end{minipage} 
\\\midrule
\mbox{Denigrates or} \mbox{intimidates}
&
\begin{minipage}[t]{2.6in}
``{\em Muslim savages brainwash their kids into hating and killing non believers, as apes and pigs, since really young}" (R)
\end{minipage}
&
Incites \mbox{violence}
&
\begin{minipage}[t]{2.6in}
``{\em \#StopIslam wipe its followers from the face of the earth}'' (T) 
\end{minipage}
\\\midrule
Diagnoses causes
&
\begin{minipage}[t]{2.6in}
``{\em the left say, look they were not refugees; the fact is that this would never happen if we would have banned them}'' (R) 
\end{minipage}
&
\mbox{Suggests a} \mbox{remedy}
&
\begin{minipage}[t]{2.6in}
``{\em we should deport Muslim scumbags and their families}" (R)
\end{minipage}
\\ \bottomrule
\end{tabular}
\caption{Example messages from Reddit (R) and Twitter (T) for some of the analyzed events, provided for illustration purposes. Messages have been (sometimes heavily) paraphrased for anonymity.}
\label{table:examples}
\end{table*}

\section{Characterizing Hate Speech}
\label{sec:characterizing}

Here, we present example themes from messages posted in the aftermath of extremist events (listed in \S\ref{sec:analysis}), 
and characterize them along four dimensions (stance, target, severity, and framing), which we then use to analyze the data.

\subsection{Exploration of Post-Event Messages}
\label{subsec:qualitative}

To understand how the content and themes of messages vary with respect to who is mentioned, what is said, and how the content is framed, 
we review messages posted after one terrorist and two Islamophobic attacks: 
{\em Manchester Arena bombing}, an Islamist terrorist attack in Manchester that targeted concert goers, killing 23 people and wounding 512 others; 
{\em Portland train attack}, carried out by a man shouting racial and anti-Muslim slurs who fatally stabbed two people and injured a third; and 
{\em Quebec City mosque shooting} that targeted worshipers, leaving 6 dead and 19 injured. 
We focus on these particular events for their overall difference in nature.
Table~\ref{table:examples} includes example messages.

\noindent {\bf Who is mentioned? } 
Naturally, many messages mentioned (directly or indirectly) Arabs, Muslims, or Islam, given how we collected our data and the focus of our study.
Yet, we also found messages mentioning the victims of the attacks, the mainstream media, political and religious groups (e.g., ``the left", ``Christians"), immigrants in general, and high-profile individuals (e.g., politicians, journalists).

\noindent {\bf What is said, and why? }  
The content of the messages ranged from blaming Arabs and Muslims for the attack, 
to providing context and defending Islam. 
Some messages made crude generalizations or included denigrating insults, while others appeared to either intimidate or incite violence.

\noindent {\bf How is the content framed? } 
According to~\citet{entman1993framing}, content may be framed according to whether it defines a problem, diagnoses its causes, makes a moral judgment, or suggests a remedy.
Several messages echoed similar points of view about what the ``problem" might be, what the causes are, and what the solutions should be. 
After the Manchester Arena Bombing, a repeated theme could be paraphrased as ``{\em if Muslims were not allowed in the country, there would be no terrorist incidents}.''
A proposed solution was e.g., ``{\em stop Islamic immigration before it is too late}." 
Some messages went further, linking the event to immigration more broadly. 
We also found posts framing the event (and what happened after) as a lesson on what Islam is, and what it stands for, e.g., ``{\em Islam only wants to kill and rape, [Q]uran is a manual of evil}."
Yet, other messages tried to push back on this framing by suggesting a counter-narrative.

\subsection{Four Dimensions of Online Hate Speech}
\label{subsec:dimensions-hate-speech}

Based on prior work and our exploration of post-event messages, we derive four main dimensions of hate and counter-hate speech: \textbf{stance}, \textbf{target}, \textbf{severity}, and \textbf{framing}.
While these are useful, we recognize these dimensions are unlikely to capture all aspects of online expressions of hate.

\noindent {\bf Stance. }
\label{subsec:stance}
~\citet{magdy2016isisisnotislam} make a distinction between online speech that {\em attacks} and blames, speech that {\em defends}, and speech that is {\em neutral} towards Islam and Muslims following a terrorist attack. 
\citet{benesch2016counterspeech} introduce a taxonomy for spontaneous expressions of counter-hate speech on social media platforms.
We adapt these categorizations to define the following stances of speech for our study: 

\begin{compactitem}[--]
\item {\em Takes a favorable stance in support of individuals, groups, or ideas: } 
defend, show solidarity, propose counter narratives, denounce, or comment on acts of hatred, or emphasize the positive traits of individuals, groups, or ideas (e.g., \#ThisIsNotIslam, \#NotInMyName); 

\item {\em Takes an unfavorable stance against individuals, groups, or ideas: } 
attack, blame, denigrate, demean, discriminate, employ negative stereotypes, seek to silence, or generally emphasizes the negative traits of an individual or group (e.g., ``kill all Muslims,'' \#RefugeesNotWelcome);  

\item {\em Commentary on negative actions or speech against individuals, groups, or ideas:  } 
comment on or characterize acts of violence, hatred, harassment, or discrimination (e.g., ``hate speech,'' ``racial slur''); and 

\item {\em Neutral, factual, or unclear if it is in support or against a person or group: } 
none of the above; report news facts or comments, describe an event, or not related to a minority or vulnerable group (e.g., ``seven injured,''  ``white van'').
\end{compactitem}

\noindent {\bf Target. }
\label{subsec:group}
Hate speech can target any minority or vulnerable group by singling out its identifying characteristics.
In the case of Muslims or Islam, these characteristics include religion, country of origin, immigration status, ethnicity, or a conflation of several or all characteristics. 
We identify the following targets of hate and counter-hate speech: 
\begin{compactitem}[--]
\item {\em Muslims and Islam}; 
\item {\em Religious groups:} unspecified, any religion except Islam;
\item {\em Arabs, Middle-Easterners, or North Africans:} descent without reference to religion; 
\item {\em Ethnic groups or groups of foreign descent:} unspecified, any foreign descent, except Arab;  
\item {\em Immigrants/refugees/foreigners in general:} without indicating a specific religion or descent; and 
\item {\em  Other groups of non-immigrants:} based on e.g., gender, sexual orientation, appearance, disability, or age. 
\end{compactitem}

\noindent {\bf Severity. }
\label{subsec:severity}
International organizations are concerned with how hate speech can lead to violent acts~\cite{izsak2015hate}.
Expressions of hate take many forms~\cite{ghanea2013intersectionality,matias2015reporting}; they can be ambiguous, and the perception of what is hateful varies between individuals~(Olteanu et al.~\citeyear{olteanu2017limits}).
Capturing such subtleties is essential to understanding how severe the repercussions of online hate speech can be;
for instance, the Jewish \emph{Anti-Defamation League} defines a ``Pyramid of Hate,'' showing how prejudice enables discrimination, which enables violence, which enables genocide.\footnote{See: \url{https://sfi.usc.edu/lessons/pyramid-hate}}
We use the following levels of severity of hate speech:

\begin{compactitem}[--]
\item {\em Promotes violence:} 
threaten with violence, incite violent acts, and intend to make the target fear for their safety (e.g., ``attack mosque,''  ``kill muslims'');

\item {\em Intimidates: } 
harass or intimidate the target, or invite others to do so, while actively seeking to cause distress (e.g., ``deport illegals,''  ``Muslims not welcomed''); 

\item {\em Offends or Discriminates: } 
defame, insult, or ridicule the target, showing bias, prejudice, or intolerance, while actively seeking to embarrass or harm the target's reputation, (e.g., ``Muslim [expletive],'' ``sand [n-word]''); 
\end{compactitem}

\noindent {\bf Framing. }
\label{subsec:framing}
\citet{kuypers2010framing} defines framing as the ``process whereby communicators, consciously or unconsciously, act to construct a point of view that encourages the facts of a given situation to be interpreted by others in a particular manner." 
\citet{benford2000framing} note that framing is critical to understand social movements and collective action;  it can also operate in different ways~\cite{entman1993framing}.
For our analysis, from test annotations we noticed that two frames were quite distinguishable in the text and complementary:\footnote{The ``makes a moral judgment'' frame is present in some form in many messages, but often supporting another frame; the ``defines a problem'' frame is rarely seen without the ``diagnoses its causes.''} 

\begin{compactitem}[--]

\item {\em Diagnoses the cause or causes for a problem} (or elements seen as possible causes): identifies what creates a problem, suggests a diagnose or disagrees with a diagnose (e.g., ``terrorists exist because they come from a place that, socially, is centuries behind''); 

\item {\em Suggests a solution or solutions for a problem} (or actions seen as possible solutions):  proposes or defends actions seen as solving or removing the problem (e.g., ``we should target the mosques and [M]uslims, this is what you need to do when at war with these [expletive]"); 

\item {\em Both diagnoses causes and suggests solutions}: if both of the above categories apply to the message.

\end{compactitem}

\noindent Terms or sentences may perform multiple of these framing functions, but they may also perform none of them~\cite{entman1993framing}.
Thus, for annotation purposes we add a catch-all category for those cases where none of these functions apply. 

\section{Methodological Framework}
\label{sec:method}

To quantify how extremist violence events affect the prevalence of various types of speech, we treat these events as interventions on observed time series.
Following existing techniques for causal inference on time series~\cite{eichler2012causal,brodersen2015inferring}, we measure this effect by comparing the behavior of an observed time series (which we refer to as {\em treated}) after an event with a counterfactual time series of its behavior had the event not taken place.  
This synthetic unobserved counterfactual time series (which we refer to as {\em control}) is modeled from several observed time series that may be correlated to the treated time series (yet not affected by the event), as we describe below.  
The causal effect is then estimated based on the differences between the treated and the control time series.
Broadly, since we model the counterfactual of the treated time series, this is a generalization of the application of the differences-in-differences techniques to time series~\cite{brodersen2015inferring}.

\begin{figure*}[tbh]
\scriptsize{\begin{filecontents*}{event_list-ordered-by-date.csv}
No, Name,Type,Language,Country,Date,Url
1,Brussels bombings,Islamist terrorism,French/Dutch,Belgium,March 22 2016,http://r.enwp.org/2016\_Brussels\_bombings
2,Orlando nightclub shooting,Islamist terrorism,English,United States,June 12 2016,http://r.enwp.org/2016\_Orlando\_nightclub\_shooting
3,Istanbul Airport attack,Islamist terrorism,Turkish,Turkey,June 28 2016,http://r.enwp.org/2016\_Ataturk\_Airport\_attack
4,Nice attack,Islamist terrorism,French,France,July 14 2016,http://r.enwp.org/2016\_Nice\_attack
5,Munich shooting,Islamophobic,German,Germany,July 22 2016,http://r.enwp.org/2016\_Munich\_shooting
6,Ohio State University attack,Islamist terrorism,English,United States,November 28 2016,http://r.enwp.org/2016\_Ohio\_State\_University\_attack
7,Berlin attack,Islamist terrorism,German,Germany,December 19 2016,http://r.enwp.org/2016\_Berlin\_attack
8,Westminster attack,Islamist terrorism,English,United Kingdom,March 22 2017,http://r.enwp.org/2017\_Westminster\_attack
9,Quebec City mosque shooting,Islamophobic,English,Canada,January 29 2017,http://r.enwp.org/Quebec\_City\_mosque\_shooting
10,Olathe Kansas shooting,Islamophobic,English,United States,February 23 2017,{http://r.enwp.org/2017\_Olathe,\_Kansas\_shooting}
11,Manchester Arena bombing,Islamist terrorism,English,United Kingdom,May 22 2017,http://r.enwp.org/2017\_Manchester\_Arena\_bombing
12,Portland train attack,Islamophobic,English,United States,May 26 2017,http://r.enwp.org/2017\_Portland\_train\_attack
13,Finsbury Park attack,Islamophobic,English,United Kingdom,June 19 2017,http://r.enwp.org/2017\_Finsbury\_Park\_attack
\end{filecontents*}
\pgfplotstabletypeset[col sep=comma, header=has colnames, 
  every head row/.append style={
        before row=\toprule,
        after row/.add={}{
        \midrule}},
   columns/No/.style={
    column name={Num.},
    string type,
    column type={@{}l}},
  columns/Date/.style={
    column name={Date},
    string type,
    column type={l}},
  columns/Name/.style={
    column name={Name},
    string type,
    column type={p{3.1cm}@{}}},
  columns/Url/.style={
    column name={URL},
    column type={l},
    string type},
  columns/Type/.style={
    column name={Type},
    column type={p{1.8cm}},
    string type},
  columns/Language/.style={
    column name={Language},
    column type={p{1.35cm}},
    string type},
  columns/Country/.style={
    column name={Country},
    column type={l@{}},
    string type},
  columns={No,Date,Name,Type,Url,Language,Country},
  every last row/.style={after row=\bottomrule},
 ]{event_list-ordered-by-date.csv}}
 \centering
\includegraphics[width=0.98\textwidth]{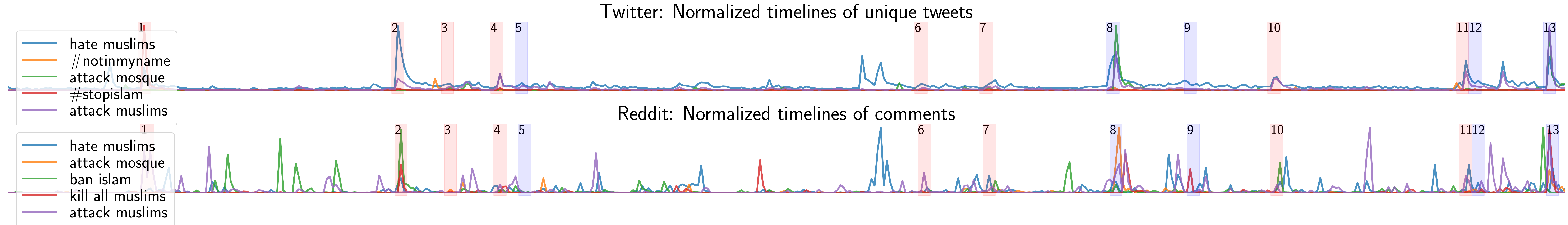}
    \caption{List of events we consider in this study (top), and examples of normalized time series corresponding to top 5 bootstrap terms by volume on both Twitter and Reddit (bottom).  Markers in the time series correspond to event numbers in the table.} 
    \label{tab:external_events}
\end{figure*}

\noindent {\bf Observed Time Series. }
We consider time series covering our 19-month observation period with a granularity of one day. 
For each of the 825 terms we have for Twitter, we experiment with three time series: one for the number of tweets, one for the number of tweets excluding re-tweets (i.e., unique after removal of ``RT @user'' prefixes), and one for the number of unique users.
Similarly, for the 1,257 terms we have for Reddit, we experiment with three time series: one for the number of posts, one for the total number of comments in these posts, and one for the total number of unique users in the post and comments.

\noindent {\bf Synthetic Control Time Series. }
A synthetic control time series is a \emph{counterfactual} that reflects behavior \emph{had the extremist violence event not taken place}. 
For each treated time series, we build a control series for 1 week following the event based on several data sources:\footnote{The 1 week observation period  allows us to observe the bulk of the effects, e.g., after a terror attack in UK, \cite{burnap2014tweeting} found that related messages were propagated for less than 2 days. }
\begin{inparaenum}[(1)]
 \item the observed series in the 11 weeks leading to the event;
 \item the observed series exactly 1 year before the event, for 12 weeks (corresponding to the 11 weeks before and 1 week after the event, but a year earlier);
 \item the observed series 23 weeks prior to the event, similarly for 12 week;\footnote{Recall that our entire dataset spans 19 months.  When information one year before is not available, we use one year later; when information 23 weeks before is not available, we use 5 weeks later.} and
 \item external information from Twitter, Reddit, and news sources.  
\end{inparaenum}

The external information includes time series whose behavior is unlikely to be affected by the events (\S\ref{sec:analysis}).
First, we use the overall volume of news on GDELT (i.e., number of daily news article URLs per day) as it does not seem to be affected by any of our events during the observation window.
Second, we use the overall number of tweets containing the word ``news'' which we also observe is not affected by any of our events (also a proxy for the overall volume of tweets).
Third, we use the overall number of Reddit posts containing the general term ``people,'' which we also observe is not affected by the events in our list (this is not the case for the series of posts in Reddit containing, e.g., the term ``breaking news'' which was affected by several of our events).

The methodology for synthesizing the control follows~\citet{brodersen2015inferring}, using a state space model to predict the counterfactual from the various sources we described above. 
However, our models are fit using maximum likelihood estimation~\cite{fultonestimating} rather than Bayesian methods like Markov chain Monte Carlo preferred by~\citet{brodersen2015inferring}.  
Our implementation uses the state space model in the \emph{UnobservedComponents} Python package to model and predict the series,
following existing Python implementations of~\citet{brodersen2015inferring}.\footnote{Python modules used: \url{http://statsmodels.org/dev/generated/statsmodels.tsa.statespace.structural.UnobservedComponents.html}; similar Python implementations \url{https://github.com/tcassou/causal_impact}, and \url{https://github.com/jamalsenouci/causalimpact}.}

\noindent {\bf Impact Estimation.  }
To estimate the effect of an event using the treatment and control time series, we compute the relative lift or drop as 
${\operatorname{rel}_{\operatorname{effect}} = 100 \times \frac{\sum t_k-c_k}{\sum c_k}}$,
where $t_k$ is the value of the treated time series at time $k$, and $c_k$ that of the control time series.  
The summations are over the days we observe after the event, seven in our case.
We focus on relative effect as it better allows for comparison across events.  
For each event, we rank terms based on the relative effect.
   
Some of our time series have intervals of low volume (particularly for Reddit) that may lead to negative-valued synthetic controls and skewed estimates of the effect.  
To address this, we add a large constant $C$ to all time series before synthesizing the control and estimating the effect.  
This transformation preserves the shape and amplitude of the impact.\footnote{We experimented with logarithmic transformations, but believe that a linear transformation leads to more interpretable confidence intervals and estimates. We arbitrarily set $C = 1000$.}

\begin{table*}[t]
\centering\scriptsize
\begin{tabular}{@{}l@{}rrrrrrrrrrrr|rrr@{}}
 & \multicolumn{4}{c}{{\bf Stance}} & \multicolumn{3}{c}{{\bf Severity}}  
 & \multicolumn{5}{c}{{\bf Target}}  & \multicolumn{3}{c}{{\bf Framing}}  \\ 
       
 & \multicolumn{1}{c}{Favor.} & \multicolumn{1}{c}{Unfavor.} 
 & \multicolumn{1}{c}{Comm.} & \multicolumn{1}{c}{Neutral}
 
 & \multicolumn{1}{@{}c}{Violent} & \multicolumn{1}{c}{Intimid.} & \multicolumn{1}{c}{Offend} 
 
 & \multicolumn{1}{@{}c}{Muslims} & \multicolumn{1}{c}{Relig.} & \multicolumn{1}{c}{Ethnic} 
 & \multicolumn{1}{c}{Immigr.} & \multicolumn{1}{c}{Non-Immigr.}  
 
 & \multicolumn{1}{@{}c}{Cause} & \multicolumn{1}{c}{Sol.} 
 & \multicolumn{1}{c}{Both}  \\ 
 
 \cmidrule(lr){2-5} \cmidrule(lr){6-8} \cmidrule(lr){9-13} \cmidrule(lr){14-16} 
\multicolumn{1}{@{}l@{}}{{\bf Reddit}}   
&  6.7\%  &  31.6\%  &  16.4\%  &  43.3\%  
&  13.8\%  &  9.8\%  &  75.8\%  
&  47.4\%  &  6.3\%  &  9.4\%  &  2.0\%  &  6.7\% 
&  64.5\%  & 2.5\%  &  9.7\%  \\ \hline

\multicolumn{1}{@{}l@{}}{{\bf Twitter}}  
&  5.1\%  &  48.3\%  &  16.7\%  &  27.0\%  
&  22.1\%  &  9.8\%  &   67.5\%  
&  40.6\%  &  4.5\%  &  11.2\%  &  3.1\%  &  8.2\% 
&  56.5\%  &  8.3\%  &  24.1\%  \\   
 \cmidrule(lr){2-5} \cmidrule(lr){6-8} \cmidrule(lr){9-13} \cmidrule(lr){14-16}  
\end{tabular}
\caption{Distribution of annotations along the entire 19-month observation period, done at the term level (except framing, done at the message level). The percentages may not add to 100\% as we omit the cases when none of the categories apply.}
 \label{tbl:annotation-results}
\end{table*}

\section{Experimental Results}
\label{sec:analysis}

In this section, we present experimental results that estimate how different types of events affect various forms of online speech.
First, we select 13 extremist violence attacks (Islamist terrorist and Islamophobic), that occurred during our full 19-month observation period.
Next, we annotate our data at query term level for stance, target, and severity, and at message level for framing, according to the hate speech taxonomy introduced in \S\ref{subsec:dimensions-hate-speech}.
Finally, we present results on various categories of hate speech across events and platforms.

\subsection{Experimental Setup}

\noindent {\bf Events Selection.  }
\label{subsec:events}
We select a set of extremist violence attacks in Western countries involving Arabs and Muslims as perpetrators or victims, and covered by international news media.
Our sources are two Wikipedia pages listing Islamist terrorist attacks and Islamophobic incidents.\footnote{\url{https://en.wikipedia.org/wiki/Islamophobic_incidents} and \url{https://en.wikipedia.org/wiki/List_of_Islamist_terrorist_attacks}}
When two events occur within the same week, we selected the one with the largest number of victims, also the most prominent in the news.
The list of events is available in Figure~\ref{tab:external_events}, where we also display the time series of top-5 bootstrap terms (\S\ref{concept:query-bootstrapping}) on Twitter and Reddit, which shows that these events cover most of the peaks in these terms for Twitter and a majority of them for Reddit.

\noindent {\bf Crowdsourced Annotations.  }
\label{subsec:annotations}
Our entire list of terms contains 1890 unique terms, which we annotate by employing crowdsource workers through the Crowdflower platform. 
We select workers from countries having a majority of native English speakers or that were affected by the events (e.g., Germany).
Except for ``framing,'' for cost and scalability purposes, we annotate each term with the most likely category the text containing them may fall under.  
For framing we annotate entire messages, as annotating at the term-level annotations does not produce reliable labels.

For each hate speech dimension and category, we provide detailed definitions and extensive examples; 
and, for each term we annotate, we show crowd workers clickable links to corresponding search results matching our queries, as returned by both social media platforms, Twitter and Reddit, as well as by two major search engines, Bing and Google. 
Following standard crowdsourcing practices, we gather at least 3 annotations per term (up to 5 when consensus was not reached), using a set of unambiguous test questions provided by the authors to catch inattentive workers, and resolving disagreements by majority voting. 
For framing, for each event we annotate samples of 5-6 messages matching the top 100 terms by relative effect, and posted around the time of the event.\footnote{Due to restrictions to the use of our main collection, to annotate samples of tweets we used the 1\% sample available at Internet Archive (https://archive.org/details/twitterstream), parsing through over 500 M tweets to locate those matching the query terms.}
To obtain the dominating frame of a term, we first determine the label of the messages it matches, and then assign by majority voting to each term the most prevalent frame, or if the ``causes" or ``solutions" frames are similarly prevalent, we assign the ``causes and solutions" frame.

Table~\ref{tbl:annotation-results} shows the overall distribution of annotations; the annotations for frame provide only an approximation based on top terms as impacted by the events in our list. 
We observed that terms marked as unfavorable represent $\approx$30\%-50\% of our query terms, and only $\approx$20\%-30\% of those are identified as particularly severe (i.e., promoting violence or intimidating); corresponding to 15\% on Twitter and 7\% on Reddit.  
Given the recall-oriented nature of our collection, this supports the observation of~\citet{faris2016understanding}, who, using a similar taxonomy, also observed that the incidence of the most severe cases of hate speech is also typically small.

\begin{figure}[tb]
  \centering
    \includegraphics[width=0.47\textwidth]{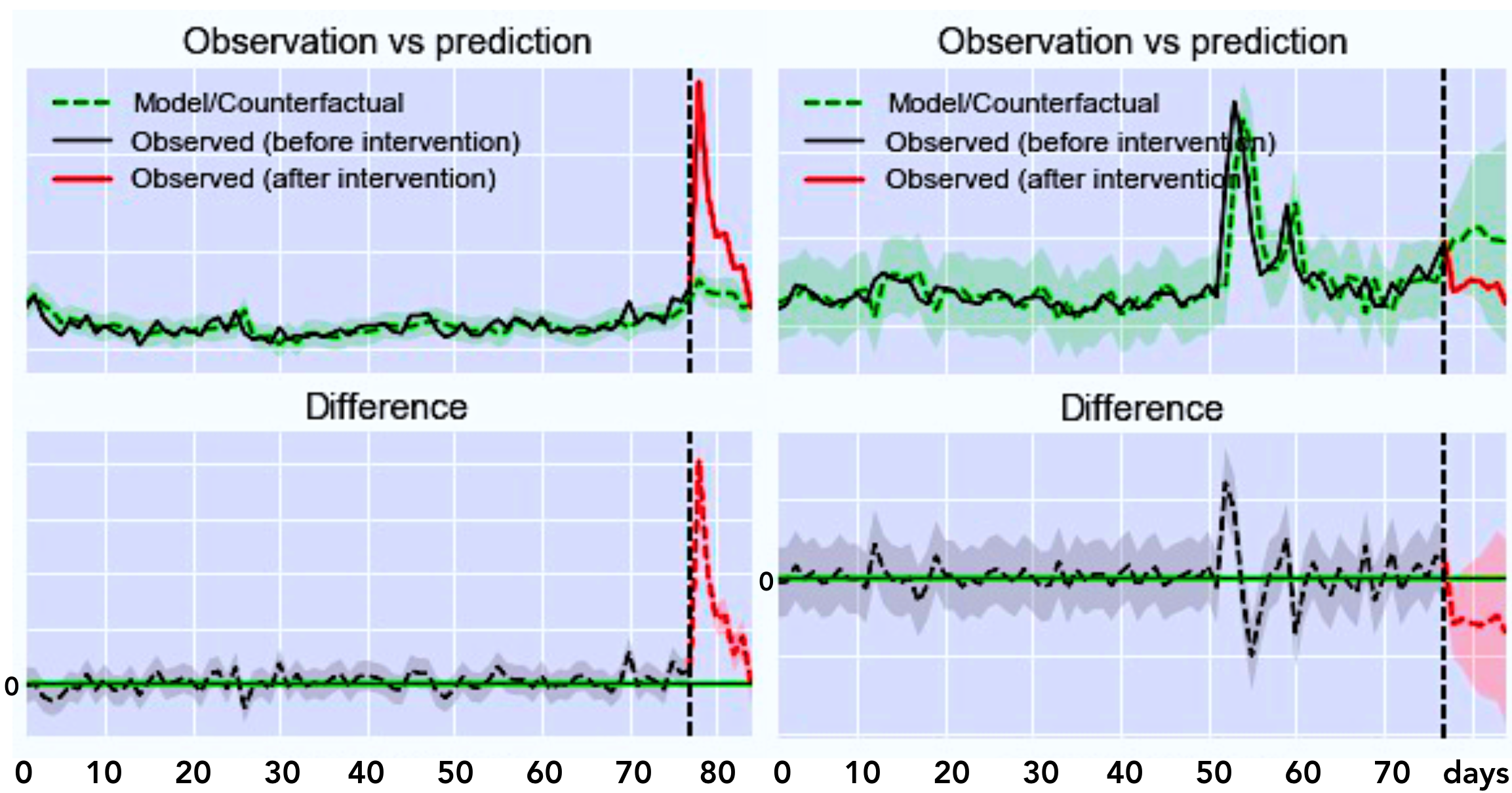}
\caption{Example of impact estimation with counterfactual predictions, for the term ``{\em evil muslims}.''
Black/red are the observed series before/after the event, green the counterfactual.
Top: time series of tweets containing the term after an Islamist terrorism attack (left: Orlando nightclub shooting) and an Islamophobic attack (right: Olathe Kansas shooting).
Bottom: differences between observed and counterfactual. 
}
\label{impact-analysis}
\end{figure}

\noindent {\bf Pre- and Post-filtering.  }
Our estimation method requires a minimum number of messages to produce a meaningful result; hence 
we filter out terms matching only a small number of messages, which we operationalize through arbitrary thresholds requiring a maximum of at least 30 users or messages per day during the event observation window. 

Figure~\ref{impact-analysis} shows an example of impact estimation on the ``{\em evil muslims}'' term, displaying the observed series, the control series, and their difference in two separate events.
In the figure, the widening confidence interval of the forecast matches the intuition that predictions become less certain as we look further into the (counterfactual) future.
In general, after applying this process, we consider there to be effect (increase or decrease) if the 90\% confidence interval of the difference between treatment and control does not include zero, which means we consider there is no effect where the 90\% confidence interval is too large or centered around zero.

\subsection{Results and Discussion}
\label{subsec:empirical-results}
In this section, we want to quantify the increase or decrease of various types of speech according to the type of event and platform.

\begin{figure}[t]
 
    \includegraphics[width=0.47\textwidth]{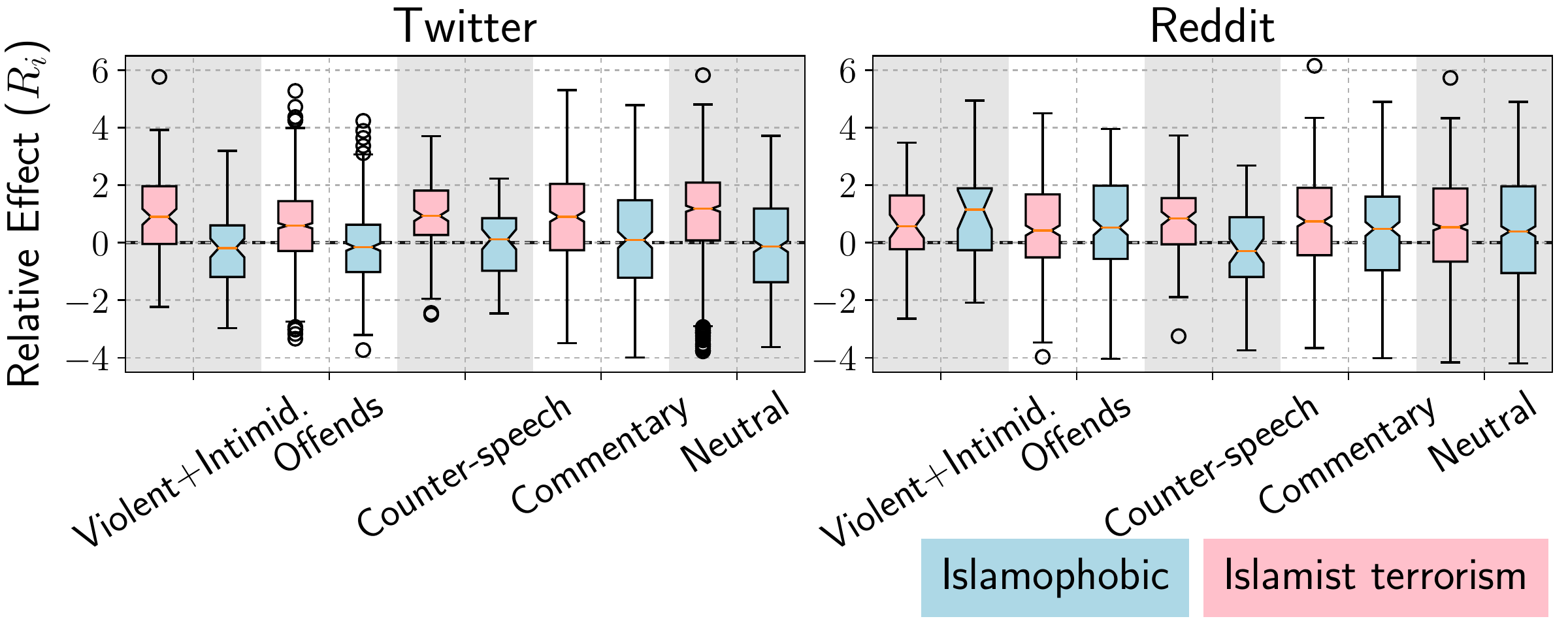}
      \caption{Distribution of impact on severity and stance across platforms (best seen in color).  
      The absolute values of the relative effect on the y-axis are log transformed such that $R_i = sign(Rel_{effect}) \times ln(abs(Rel_{effect}))$.
 } \centering
 \label{plot:severity}
\end{figure}

\begin{figure}[t]
  \centering 
    \includegraphics[width=0.47\textwidth]{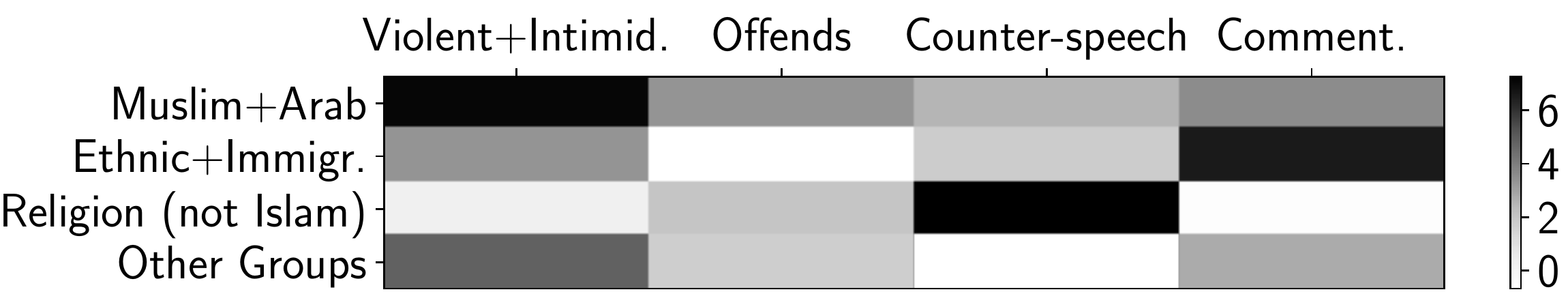}
    \caption{Aggregated variations in the mean relative effect across targets and types of speech on both platforms.}
    \label{plot:summary} 
\end{figure}
 
\noindent {\bf Do distinct types of events impact hate and counter-hate speech in different ways?}
Prior work found an increase in {\em hate speech} targeting Muslims following Islamist terrorist attacks~\cite{faris2016understanding},
and our results agree (Twitter~(T): +3.0, 95\%CI [1.7, 4.4], Reddit~(R): +2.9, 95\%CI [2.4, 3.3]).\footnote{Throughout the evaluation, these values indicate point estimates for the mean relative effect for each referenced category.}
Looking at the intersection of high severity categories (``promotes violence'' and ``intimidates'') with the target categories for Muslims and Arabs, we estimate an increase in the relative effects across events in both platforms (T: +10.1, 95\%CI [1.4, 18.9], R: +6.2, 95\%CI [3.9, 8.4]); also higher than the less severe category (offends or discriminates), with one exception, the 2016 Istanbul Airport attack.

The question is whether Islamophobic attacks elicit a similar, consistent reaction across platforms and events.
The answer seems to be no:
for instance, we only observe this pattern after one Islamophobic attack (the 2016 Finsbury Park attack), while after the 2017 Olathe Kansas shooting we estimate a decrease in high severity terms in both platforms.
This observation is also supported at an aggregate level by Figure~\ref{plot:severity} (per-event figures omitted for brevity).

Similarly, our estimates indicate an overall increase in {\em counter-hate speech} terms following Islamist terrorist attacks (T: +1.8,  95\%CI [0.7, 3.0], R: +2.9, 95\%CI [2.4, 3.4]), but not after Islamophobic attacks. 
This effect of Islamist terror attacks on counter-speech is consistent with \citet{magdy2016isisisnotislam} who noticed a notable number of messages defending Muslims and Islam following the 2015 Islamist terror attack in Paris.

\begin{figure}[tb]
  \centering
    \includegraphics[width=0.47\textwidth]{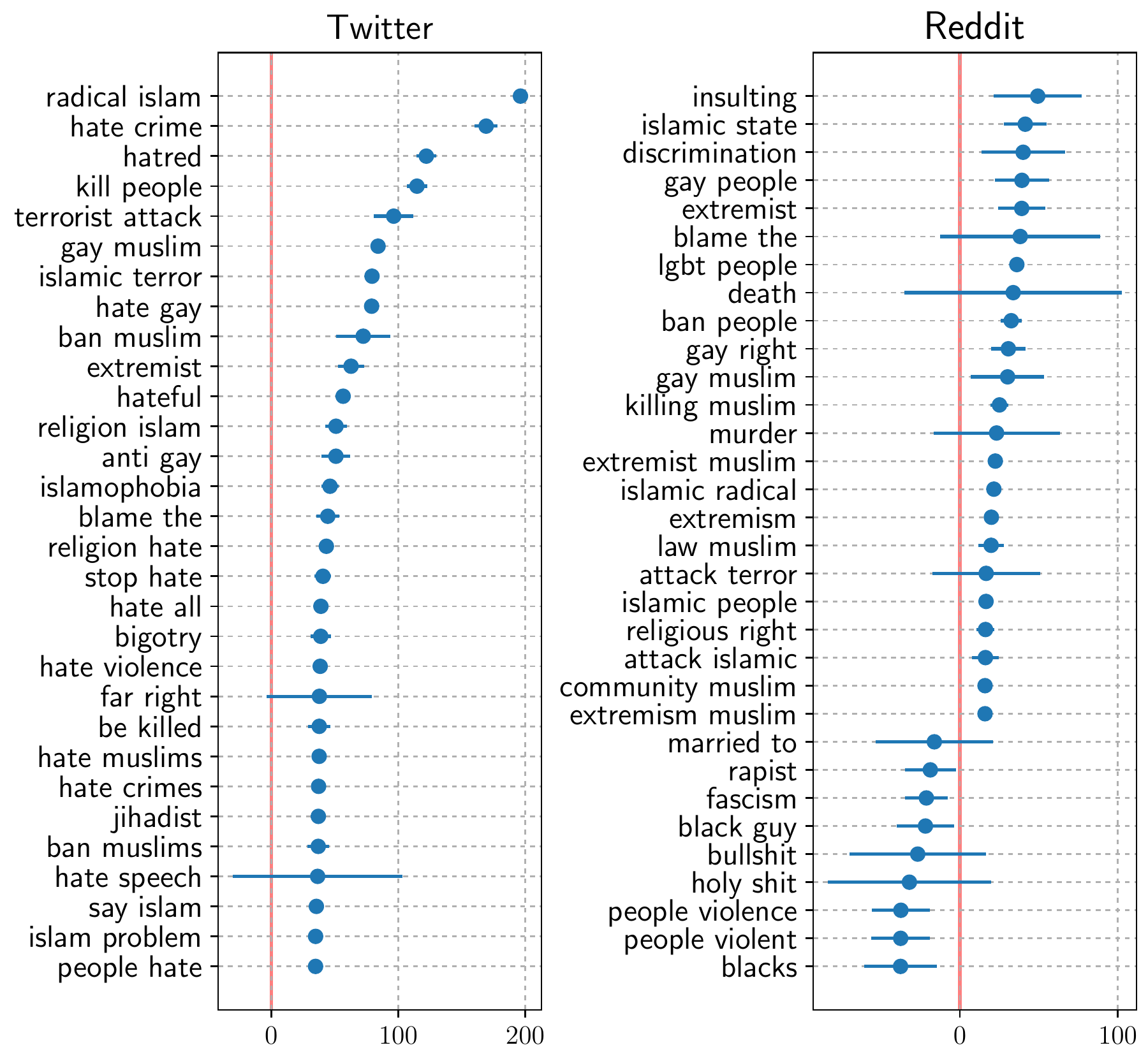}
\caption{Increase and decrease of top terms after the 2016 Orlando nightclub shooting, where a gay venue was attacked by a self-identified ``Islamic soldier.''}
\label{plot:impact-dist}
\end{figure}

\noindent {\bf Are these events more likely to lead to an increase in a specific type of speech?}
Figure~\ref{plot:summary} suggests that, on average, there is a higher increase in speech that both promotes violence or intimidates and focuses on Muslims and Arabs following extremist violence events;  
while there is an increase in counter-hate speech related to religion but not specifically mentioning Islam, e.g., focusing on religious tolerance or on positive aspects of religion in general (T: +2.8, 95\%CI [2.1, 3.6], R: +7.8, 95\%CI [1.9, 14.5]).

At the event-level, Figure~\ref{plot:impact-dist} showcases an example of a complex interplay between hate and counter-hate speech terms in reference to different groups after the 2016 Orlando nightclub shooting.  
This was not only a deadly Islamist terrorist incident, but also the deadliest homophobic attack in the U.S., which means it was very prominently covered in media.
It triggered a substantial increase in terms referring to both Islam and the gay/LGBT community. 
In general, our observations agree with an increase in mentions of Muslims, Islam, or Arabs, after Islamist terror attacks; but not after Islamophobic attacks (figures omitted for brevity).

\begin{table}[tb]
\centering\scriptsize
\begin{tabular}{@{}lrrr|rrr@{}}
 &\multicolumn{3}{c}{{\bf Islamist terrorism}} & \multicolumn{3}{c}{{\bf Islamophobic}}  \\
 & \multicolumn{1}{c}{Cause} & \multicolumn{1}{c}{Solution} 
 & \multicolumn{1}{c}{Both} 
 & \multicolumn{1}{|c}{Cause} & \multicolumn{1}{c}{Solution} 
 & \multicolumn{1}{c}{Both} \\ \midrule
 \multicolumn{1}{@{}l}{{\bf Reddit}}  &  64.4\% &  2.6\%  &    7.5\% &  64.8\% &  2.5\%  &  13.3\% \\ 
  \multicolumn{1}{@{}l}{{\bf Twitter}}       &  54.4\%  &  9.2\%  &  25.2\%  &  59.6\%  &  6.7\%  &  22.4\%  \\ 
\bottomrule
\end{tabular}
\caption{Framing distribution for top terms (percentages do not add to 100\% as we omit the ``Does not apply" category).}
\label{tbl:framing}
\end{table}

\noindent {\bf Are there differences in how hate speech is shaped by the events across platforms?}
For Twitter, Figure~\ref{plot:severity} suggests an increase for the high-severity categories (``promotes violence'' and ``intimidates'') after Islamic terrorist attacks, but not after Islamophobic attacks.
In contrast, for Reddit this distinction is absent, as we see an overall increase after both Islamist terrorist and Islamophobic attacks.

Another aspect in which we see differences between Twitter and Reddit is in terms of the framing of messages, particularly with respect to messages including a ``solution or something seen as a solution.''
In general, the terms that tend to increase the most in this frame call for banning or deporting immigrants/Muslims/Arabs, or waging war against Islam and/or Arabs.
As shown in the ``solution'' and ``both'' columns in Table~\ref{tbl:framing}, this fraction is more prevalent for Twitter among the top 100 most impacted terms (about 34\% in Islamist terrorist attacks, about 29\% in Islamophobic attacks) than for Reddit (about 10\% and 16\% respectively).

\section{Conclusions}
\label{sec:conclusion}

Measuring the effect of external events on hate speech on social media is a challenging task, which needs to be approached with an appropriate methodology (causal inference in our case), and requires a combination of automated processes and manual annotations that balances the needs of large-scale analysis with a finite human annotation budget.

We used data from two social media sites, and from two classes of events (Islamist terrorism and Islamophobic attacks), performing a comparison of observed time series for various classes of online hate speech during such events, with counterfactual series that approximate their behavior had those events not taken place.
This allows us to make more precise observations about the evolution of hate (and counter-hate) speech of various classes.
Our methodology and observations provide a blueprint for better monitoring hate speech online, with particular attention to the relation between calls for violence online and deadly extremist attacks.
Additionally, as we estimate increases in counter-hate speech during these attacks, social media platforms could intervene by boosting its visibility.

\noindent {\bf Future Work and Limitations.}
\label{limitations}
We hope that the evidence of variations in hate speech following certain events will lead to further research to understand why it happens, who it happens to, and what other qualities of an event may explain these variations; as well as research that delves into the source of the differences we observed across platforms. 

Further, while our data collection is designed to maximize recall, aiming to provide a good coverage across several categorization dimensions,
our bootstrap list of terms can still lead to bias in what gets included in our collections or not. 
The reliance on query-level annotations may as well introduce noise and biases due to ambiguous uses of some of the terms. 
We focused on English and only 13 events in the ``West,'' yet future work includes explorations into how our observations may translate to other regions, languages, and type of events. 
Our frame analysis is also only a first stab at how hateful content is framed after extremist attacks; more in-depth analyses are needed.  

Finally, our analysis is retrospective, and harmful content is actively deleted by many social media platforms when reported~\cite{matias2015reporting}, which can result in incomplete data collections. 
As a result, we are more confident in results indicating an increase in certain types of speech, than on those indicating a decrease.

\noindent {\bf Reproducibility. }
The list of our query terms, several example time series, and the detailed instructions used in the crowdsourcing tasks, are available for research purposes at \url{https://github.com/sajao/EventsImpactOnHateSpeech}.

\noindent {\bf Acknowledgments. }
We thank Miguel Luengo-Oroz for early discussions and anonymous reviewers for their comments.
This work was conducted under the auspices of the IBM Science for Social Good initiative.  
C. Castillo is partially funded by La Caixa project LCF/PR/PR16/11110009. 

\bibliographystyle{aaai}
\fontsize{9.0pt}{10.0pt}
\selectfont
\bibliography{ref} 

\end{document}